\documentclass[%
aip,
amsmath,amssymb,
reprint,%
]{revtex4-1}

\usepackage[utf8]{inputenc} 
\usepackage[T1]{fontenc}    
\usepackage{hyperref}       
\usepackage{url}            
\usepackage{booktabs}       
\usepackage{amsfonts}       
\usepackage{nicefrac}       
\usepackage{microtype}      
\usepackage{latexsym,exscale,amssymb,amsmath,amsfonts,amssymb, amscd, amstext, mathrsfs, bigints, relsize}	
\usepackage{graphicx}						
\usepackage{relsize}
\usepackage{eso-pic}						
\usepackage{float}							
\usepackage{wrapfig}						
\usepackage{siunitx}
\DeclareSIUnit\inverse{^{-1}}
\DeclareSIUnit\inversesquare{^{-2}}
\DeclareSIUnit\year{yr}
\DeclareSIUnit\carbon{GtC}
\DeclareSIUnit\humans{H}
\DeclareSIUnit\dollar{\$}
\DeclareSIUnit\bits{bits}

\begin{document}

\preprint{AIP/123-QED}


\title{Deep reinforcement learning in World-Earth system models to discover sustainable management strategies}

\author{Felix M. Strnad}
\email{strnad@pik-potsdam.de}
\affiliation{Potsdam Institute for Climate Impact Research, FutureLab on Game Theory and Networks of Interacting Agents, Research Department 4: Complexity Science, 14473 Potsdam, Germany}%
\affiliation{Department of Physics, University of Göttingen, 37077 Göttingen, Germany}
\author{Wolfram Barfuss}%
\affiliation{Potsdam Institute for Climate Impact Research, FutureLab on Earth Resilience in the Anthropocene, Research Department 1: Earth System Analysis, 14473 Potsdam, Germany}%
\affiliation{Max Planck Institute for Mathematics in the Sciences, 04103 Leipzig, Germany}

\author{Jonathan F. Donges}
\affiliation{Potsdam Institute for Climate Impact Research, FutureLab on Earth Resilience in the Anthropocene, Research Department 1: Earth System Analysis, 14473 Potsdam, Germany}%
\affiliation{Stockholm Resilience Center, Stockholm University, 104 05 Stockholm, Sweden}

\author{Jobst Heitzig}
\affiliation{Potsdam Institute for Climate Impact Research, FutureLab on Game Theory and Networks of Interacting Agents, Research Department 4: Complexity Science, 14473 Potsdam, Germany}%

\date{\today}%

\begin{abstract}
     Increasingly complex, non-linear World-Earth system models are used for describing the dynamics of the biophysical Earth system and the socio-economic and socio-cultural World of human societies and their interactions. Identifying pathways towards a sustainable future in these models for informing policy makers and the wider public, e.g. pathways leading to a robust mitigation of dangerous anthropogenic climate change, is a challenging and widely investigated task in the field of climate research and broader Earth system science. This problem is particularly difficult when constraints on avoiding transgressions of planetary boundaries and social foundations need to be taken into account.
    In this work, we propose to combine recently developed machine learning techniques, namely deep reinforcement learning (DRL), with classical analysis of trajectories in the World-Earth system. Based on the concept of the agent-environment interface, we develop an agent that is generally able to act and learn in variable manageable environment models of the Earth system.
    We demonstrate the potential of our framework by applying DRL algorithms to two stylized World-Earth system models. Conceptually, we explore thereby the feasibility of finding novel global governance policies leading into a safe and just operating space constrained by certain planetary and socio-economic boundaries. 
    The artificially intelligent agent learns that the timing of a specific mix of taxing carbon emissions and subsidies on renewables is of crucial relevance for finding World-Earth system trajectories that are sustainable on the long term.
\end{abstract}

\maketitle

\begin{quotation}
    We propose a framework for using deep reinforcement learning (DRL) as an approach to extend the field of Earth system analysis by a new method. We build our framework upon the agent-environment interface concept. The agent can apply management options to models of the Earth system as the environment of interest and learn by rewards provided by the environment. We train our agent with a deep Q-neural network extended by current state-of-the-art algorithms. We find that the agent is able to learn novel, previously undiscovered policies that navigate the system into sustainable regions in two exemplary conceptual models of the World-Earth system. 
    
\end{quotation}

\section{Introduction}

Efforts invested in identifying pathways towards global sustainability need to account for critical feedback loops between the socio-economic and socio-cultural World and the biophysical Earth system \cite{Schellnhuber1999,Donges2017}. 
These pathways may require novel, yet undiscovered multi-level policies, from local to the global scale, for the governance of this coupled World-Earth system leading towards a safe and just operating space  \cite{Rockstroem2009, Rockstroem2009b}. 
Striving for a safe and just operating space, policymakers of the United Nations agreed on global political cooperation for a sustainable future at the resolution of the 17 Sustainable Development Goals (SDG) \cite{UNSDG2015} and the adoption of the Paris Agreement on Climate Change \cite{COP2015}.
The safe and just operating space is based on a set of biophysical planetary boundaries (defined on dimensions such as climate change or biosphere integrity loss) as they are formulated by \citeauthor{Rockstroem2009} in \cite{ Rockstroem2009,Rockstroem2009b, Anderies2013, Steffen2015} extended by social foundations (e.g. poverty alleviation) by \citeauthor{Raworth2012} \cite{Raworth2012}. If respected together, staying within these boundaries is seen as a prerequisite to ensuring sustainable human development. The field of Earth system modeling develops computer models to show possible pathways towards a sustainable future. However, the identification and characterization of concrete trajectories within the planetary boundaries and above social foundations remains a problem requiring ongoing research efforts \cite{Rogelj2018, Steffen2018}.

In this paper, we consider this problem on a globally aggregated level assuming the following basic structure: An abstract single decision-maker interacts with a dynamical, in most cases non-linear environment to find sustainable trajectories within certain boundaries.
The field of \textit{Integrated Assessment Modeling} (IAM) addresses this issue via optimizing a social welfare function in order to estimate the design of sustainable management strategies  \cite{Mueller-Hansen2017}. IAM models integrate data and knowledge from established climate models \cite{Kelly1999, PahlWostl2000}. To identify pathways in IAM, numerical solvers such as GAMS \cite{Bussieck2004} are frequently used. However, these IAM models are highly dependent on the choice of the target function of the optimization. In many cases, this choice may not be obvious and depends on the IAM developers \cite{Pindyck2017}. 


As another approach, \textit{optimal control theory} (OCT) can be used to solve problems where dynamical systems are supposed to stay within certain constraints. In these systems, OCT tries to find an optimal choice for some control variable by optimizing a specific objective function \cite{Kamien2012}. Applied to Earth system models, the focus has been set on the design of climate regulators and their impact on climate modification \cite{Liang2008, Botta2018}. \textit{Viability theory} (VT) as a subfield of OCT can be stated as an example. In this field, such problems of identifying trajectories are typically addressed by methods that rely on a discretization of the state space, followed by the application of local linear approximations \cite{Deffuant2011}. It is however not well applicable in systems with more than just a small number of variables due to the curse of dimensionality \cite{Kittel2017a}. 

The use of \textit{reinforcement learning} (RL) \cite{Sutton1998} can also be considered as a possible approach for intelligent decision making within World-Earth system models\cite{Osten2017}.  
It is designed for finding optimal policy strategies as well. But in contrast to the previously presented approaches, RL does not detect solutions based on numerically solving an optimization problem, but by a dynamic search process via exploration and exploitation of past experiences, guided by a reward function. However, tabular methods, which are mainly used for classical RL solutions, cannot be straightforwardly applied to the systems of interest here, due to the continuous state spaces that we mostly find in World-Earth system models. 

The common point of all the presented methods outlined above is that they reach their limits as the complexity of the environments increases. 
However, \textit{deep reinforcement learning} (DRL) \cite{Mnih2015} algorithms have been shown to detect solutions in other highly complex environments surprisingly well \cite{Mnih2013,Mnih2015}. In this paper, we propose using DRL as a novel approach for Earth system analysis.
Even though first successful reinforcement learning experiments by using neural networks as nonlinear function approximators were reported already in 1995 \cite{Tesauro1995}, the breakthrough of DRL was achieved only in 2013 \cite{Mnih2013, Mnih2015}. Since then, DRL algorithms have become increasingly popular in the field of Artificial Intelligence \cite{Arulkumaran2017,Li2018}. The key to success of this approach lies in the combination of Q-learning\cite{Watkins1989}, neural networks \cite{LeCun2015} and experience replay \cite{Lin1993b} which has been shown to learn policies up to a super-human performance in a variety of different environments \cite{Mnih2013,Mnih2015}. Often DRL-applications come up with unexpected and novel solutions \cite{Silver2016, Silver2018}. Many extensions have been proposed addressing both speed and efficiency \cite{Hessel2018}. Due to its general applicability to various environments, DRL is used in a wide range of different fields, e.g. resources management in computer clusters \cite{Mao2016}, optimization of chemical reactions \cite{Zhou2017}, playing abstract strategy games like chess and Go \cite{Silver2016, Silver2018}, autonomous driving \cite{Lillicrap2016}, and in particular robotics \cite{Levine2016,  Zhu2017,Gu2017, Haarnoja2018}. 

Due to the wide applicability of DRL, we propose a framework that uses DRL as a tool that is both robust and easy to use at the same time to identify and classify trajectories in Earth system models effectively. As a proof of concept, we use our DRL framework within various stylized World-Earth system models \cite{Donges2017, Donges2018}. These models are designed to investigate the coevolutionary dynamics of humans and nature in the Anthropocene. Some first applications of reinforcement learning methods within resource use models have been carried out \cite{Arthur1991,Lindkvist2014, Lindkvist2017}, but as far as we know, there are no approaches yet applying DRL to Earth system models. 
We believe this approach will open so far unused possibilities to discover so far unknown management strategies that keep the Earth system within planetary boundaries, while, at the same time, respecting social foundations of the world's societies. 
Recently, various ways of how to tackle problems related to anthropogenic climate change by using machine learning techniques have been outlined \cite{Rolnick2019}. 
Our work proposes a novel strand to this list.

\section{Methods}
This work uses the agent-environment interface \cite{Sutton1998} as a formal mathematical framework which allows for making a fruitful connection between reinforcement learning and the modeling of social-ecological systems, as it was, e.g., proposed by \citeauthor{Barfuss2019} \cite{Barfuss2019}. In the case of a single agent as studied here, RL problems are based on the concept of Markov decision processes (MDPs) \cite{Sutton1998}. Therefore, we will provide a brief introduction to MDPs, followed by a description of how we included the learning process by using neural networks. We will further give a short overview of possible extensions and conclude this section by outlining how we translate Earth system models into the formal framework of an MDP.

\subsection{Markov Decision Processes} \label{sec:mdp}
RL is designed for problems where an agent observing the environment output consisting of a state and a scalar reward signal is acting upon this observation \cite{Sutton1998}. Formally, this interaction is described by a so-called Markov decision process (MDP) \cite{Wiering2012}. At each step $ t $ the environment is in a certain state $ s_t \in \mathcal{S} $, where $ \mathcal{S} $ describes the set of all possible states.  The agent chooses an action $ a_t $ from a given finite action set $ a_t \in \mathcal{A} $. Environmental dynamics are now determined by the transition probability $ T(s'| a,s)=P(s_{t+1}=s'|s_t=s,a=a_t) $ which does not depend on $t$ explicitly. When for a given action $ a $ the environment transits from state $ s $ to $ s' $, the agent receives an immediate numerical value $ r_t $, called the \textit{reward}, that generally depends on the state $s$ and action $a$. The tuple $(s,a,r,s')$ defines the MDP. The agent chooses its action according to its behavior policy $ \pi $ which maps state $ s $ to a probability distribution over all actions $ a\in \mathcal{A} $, expressed as $ \pi(s,a)=P(a|s) $. 

\subsection{Deep Reinforcement Learning} \label{sec:drl}
Every decision the agent takes is followed by a reward it gets from the environment. In all types RL algorithms, the goal of an agent is to maximize its exponentially discounted sum of future rewards \cite{Sutton1998}, called the gain $ G_t= r_t + \gamma r_{t+1} + \gamma^2 r_{t+2} + \cdots = \sum_{k=0}^{\infty} \gamma^k r_{t+k}$, where the discount factor $ \gamma \in [0,1] $ expresses how much the agent cares for future rewards. This lets us define a state-action value function $ Q_{\pi} $ quantifying the value of a state $ s $,  given that the agent applies action $ a $, as the expected return, following a given policy $ \pi $, $ Q_{\pi}(s) = \mathbb{E}_{\pi}[G_t | s_t=s, a_t=a] $. The average $\mathbb{E}_{\pi} $ can be understood as the sum over all actions for a policy $ \pi $ times the sum over all possible state transitions to $ s' $. Inserting the gain $ G_t $ yields the \textit{Bellman equation} \cite{Bellman1957}, 
\begin{equation}
Q_{\pi}(s,a) = \mathbb{E}_{\pi}\left[r_t + \gamma Q_{\pi}(s_{t+1}, a_{t+1}) | s_{t}=s, a_t=a\right].
\end{equation}
The best possible solution of an MDP is the optimal state-action-value function $ Q^*(s,a) $ which is the maximum state-action value function over all policies  
\begin{equation}
Q^*(s,a) = \max_{\pi}Q_{\pi}(s,a).
\end{equation}
The problem of maximizing the expected discounted reward sum $G_t$ is transformed to find the optimal state-action value function $Q^*$.  The optimal value function allows the following consideration. If for all possible actions $ a' $ for the next time step $ s'=s_{t+1} $ the value of $ Q^*(s',a') $ was known, then the optimal strategy would be to choose that $ a' \in \mathcal{A}$ resulting in the highest value of $ Q^*(s',a') $. This identity is known as the \textit{Bellman Optimality Equation} \cite{Sutton1998},
\begin{equation}
Q^*(s,a) = \mathbb{E}_{T}\left[ r + \gamma \max_{a'\in \mathcal{A}} Q^*(s', a')  \right]. \label{eq:BOE}
\end{equation}
$\mathbb{E}_T $ averages over all possible state transitions, given by the transition probability $ T $. 
The task is now to find a way to estimate the optimal action-value function $ Q^*(s,a) $. Estimating the state-action value function by performing rollouts on the environment are called \textit{model-free}. \citeauthor{Mnih2015} \cite{Mnih2013,Mnih2015} address this issue with the combination of Q-learning, deep neural networks and experience replay successfully, called \textit{Deep Q-learning} (DQL). Briefly, we will provide an overview of their approach.

\paragraph{Q-learning}
Q-learning is a specific type of RL which converges to the optimal solution.
 In Q-learning we use the function $ Q(s, a) $ representing the state-action value when performing action $ a $ in state $ s $.
The temporal difference error of expected value $Q(s,a)$ and experienced value $r + Q(s',a')$ is used to estimate the current value of the state \cite{Sutton1998}. 
It is used to incrementally estimate Q-values for actions, based on an iteratively updated $ Q $-value function \cite{Watkins1989},
\begin{equation}
Q_{i+1}(s_t,a)=r_t + \gamma \max_{a'\in \mathcal{A}} Q_i(s_{t+1}, a'). \label{eq:qlearning}    
\end{equation}
Action selection when acting in the environment is usually made with an $ \epsilon $-greedy policy, i.e., with probability $ \epsilon \in [0,1] $ the action $\mathrm{argmax}_a Q(s,a)$ is used and with probability $1- \epsilon $ a random action is used. Here, the parameter $\epsilon$ regulates this exploration-exploitation trade-off. Q-learning is an \textit{off-policy} algorithm, i.e. to estimate the current state-action value the agent uses the maximum state-action value of the next state, regardless which action is actually chosen there. Still, one can prove that for $ i\rightarrow \infty $ this algorithm will converge to the optimal action value function $ Q(s,a) \rightarrow Q^*(s,a) $ \cite{Sutton1998}.

\paragraph{Deep Q-Networks}
In practice, this convergence is only applicable in state spaces with a small number of states. However, continuous state spaces make it impossible to learn state-action pairs independently \cite{Sutton1998}. Using multi-layered neural networks as function approximators, $ Q(s,a, \theta)\approx Q^*(s,a) $, called Deep Q-Networks (DQN), is a possibility to overcome this issue \cite{Mnih2015}. 
As target function $ Y_t $ one can use different RL variants \cite{Sutton1998}. Here the Q-learning update from equation (\ref{eq:qlearning}) is adjusted by setting $ Y_t(s_t,a_t, \theta)=r_t + \gamma \max_{a'\in \mathcal{A}} Q(s_{t+1}, a', \theta)$.
The parameters (i.e., the weights) $ \theta_i $ of the neural network are optimized by gradient descent to minimize the loss $\mathcal{L}_i (\theta_i) $ at iteration $ i $ between the target and the current $Q$ value via
\begin{align}
\mathcal{L}_i (\theta_i)        &= \left( Y_t(\theta^-_i) - Q(s_t,a_t, \theta_i) \right)^2,\label{eq:loss_function} \\ 
\nabla_{\theta_i} \mathcal{L}_i &= \left( Y_t(\theta^-_i) - Q(s_t,a_t, \theta_i)\right) 
\nabla_{\theta_i}Q(s_t, a, \theta_i), \\
\theta_{i+1} &= \theta_i +\alpha \nabla_{\theta_i} \mathcal{L}_i.
\end{align}
The parameter $\alpha$ describes the learning rate of the network.
To account for a more stable learning a second network with parameters $ \theta_t^- $ is used. This network is a copy of the first one but is frozen in time for $\tau_{target}$ iteration steps. It is used as the target function $Y_t(s,a,\theta^{-}) $ in equation (\ref{eq:loss_function}). The fixed $ Q $-values of $Y_t(s,a,\theta^{-}$  make it possible, that the optimization process converges to a stable target \cite{Mnih2015}.
The target network is updated every $ \tau_{target} $ iteration steps by copying the parameters from the current network: $ \theta^- \leftarrow \theta $.

\paragraph{Experience replay}
Instead of learning from state-action pairs as they occur during simulation, updates for the state action value function $Q(s,a,\theta$) are applied on samples (called Mini-Batches) randomly drawn from a replay memory \--- typically a large table of stored observations, that are collected during the training process. This separates the learning process itself from gaining experience \cite{Lin1993b} which breaks the similarity of subsequent training samples and leads consequently to more stable learning \cite{Mnih2015}.

\subsection{Extensions to DQN} \label{sec:extensionsDQN}
After the convincing performance of the DQN network presented by \citeauthor{Mnih2015} \cite{Mnih2015} the algorithm has been further developed and significant improvements regarding stability and learning speed could be achieved. 
By using double Q-learning \cite{VanHasselt2016} harmful overestimation of the $ Q $ values \cite{Hasselt2010} can be reduced. With the introduction of dueling network architectures \cite{Wang2015} the value of a state and the advantage of taking a certain action at that state could be decoupled. 
Furthermore, the distributional DQL algorithm by \citeauthor{Bellemare2017} \cite{Bellemare2017} addresses the issue that the value of future rewards is restricted to the expected return (i.e., to the $Q$ function) and replaces it with a distribution of $Q$-values per action. 
These improvements are often combined with prioritized experience replay \cite{Schaul2015} that privileges experiences from which the agent can learn more. In \cite{Hessel2018} \citeauthor{Hessel2018} compare and combine improvements  to a new state-of-the-art DQL algorithm, they called \textit{Rainbow}, which we will also use in this paper.

\subsection{Agent-Environment Interface} \label{sec:aei}
In this work, we transfer the theoretical framework of an MDP to concrete applications in Earth system dynamics by using the agent-environment interface. In this context, the concept of the agent is solely defined by its action set. 
The action set can be regarded as a collection of possible measures the international community could use to influence the system's trajectory. The agent uses a DRL algorithm outlined in section \ref{sec:drl}. 
Concerning the detection of sustainable governance policies, we are mostly interested in the final outcomes the agent has learned rather than in letting the agent make real-world decisions later on. 
To assess the feasibility of finding sustainable policies, we also investigate the learning process. 
In this work, we intend to test our framework in the context of Earth system models. We focus on a particular type of Earth system models, which has been termed ``World-Earth models'' \cite{Donges2017, Donges2018}. In World-Earth modeling, one tries to capture the coevolving dynamics between biophysical dynamics of the Earth system on the one hand and on the other hand the social and economic dynamics of the World community. 
Since optimizing welfare may lead to policies which are neither sustainable nor safe \cite{Barfuss2018}, we are interested in governance policies whose resulting trajectories stay within certain \textit{``sustainability boundaries''} of the state space. These include both planetary and socio-economic boundaries. We set up the environments based on \citeauthor{Kittel2017a} \cite{Kittel2017a} and \citeauthor{Nitzbon2017}\cite{Nitzbon2017}, both using deterministic nonlinear World-Earth models including planetary boundaries and social foundations. The dynamics are described by a set of coupled autonomous differential equations that define a continuous state space. In our setting, time is discretized in integration steps $ \text{d}t $. At each $ n $-th step, the environment's dynamics are numerically solved (i.e., integrated) for a single timestep $ \text{d}t_n= t_{n}-t_{n-1} $. Therefore, the environments fulfill the Markov property of the MDP. 

\begin{figure*}
    \centering
    \includegraphics[width=1.\linewidth]{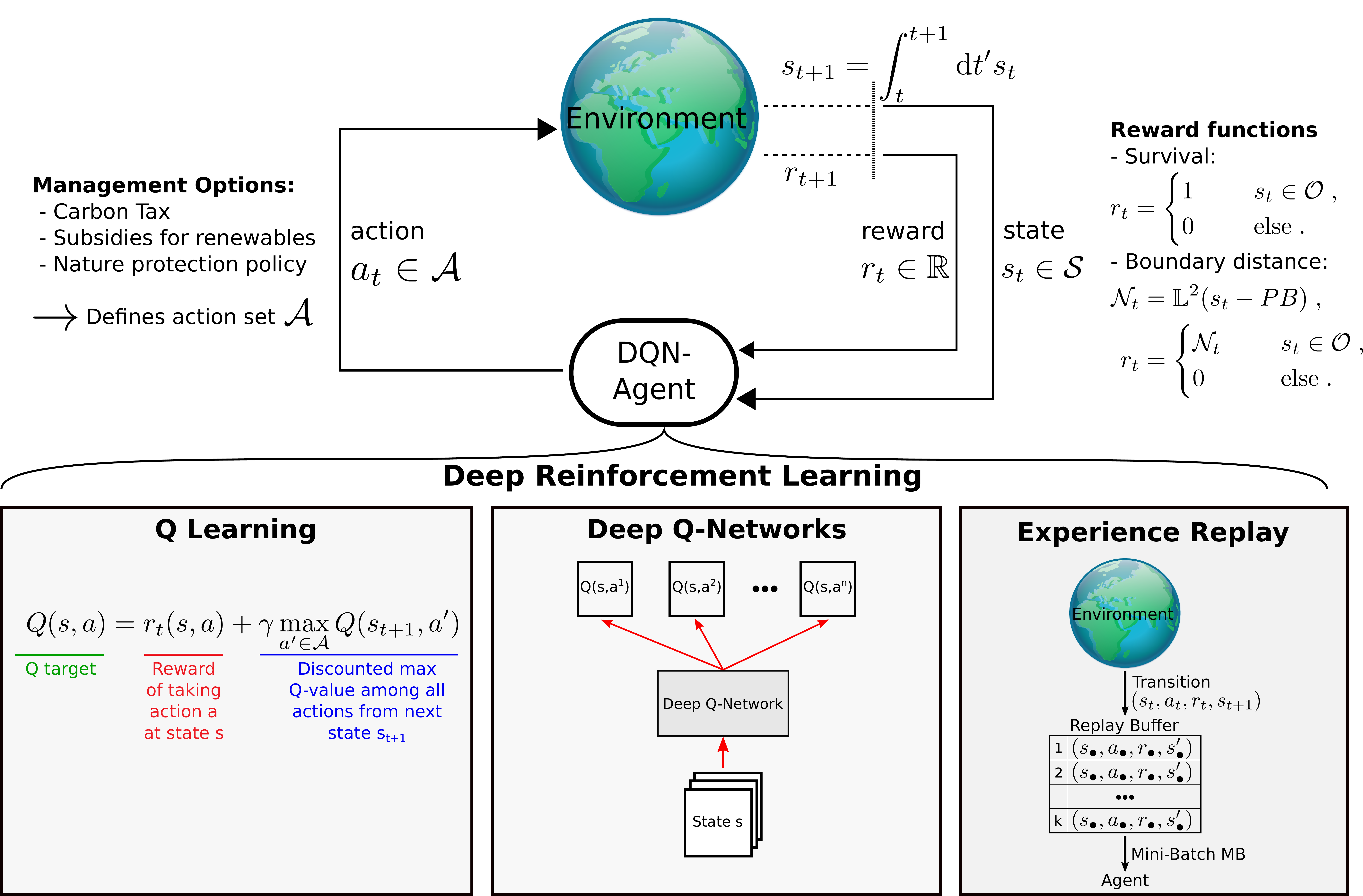}
    \caption{Using the agent-environment interface \cite{Sutton1998} for analyzing World-Earth models via deep reinforcement learning (DRL) techniques. The environment is in a certain state $s_t$, based on that state the agent chooses an action $a_t$. The environment responds with a next state $s_{t+1}$ and a reward $r_{t+1}$. The dynamics of the environment for every timestep $dt$ are numerically integrated. We interpret the action set $\mathcal{A}$ as a set of management options for the Earth system and propose different reward functions. $\mathcal{O}$ describes the set of states that are within the planetary boundaries (PB).
    The learning agent is implemented to use DRL \cite{Mnih2013, Mnih2015} using Q-Learning \cite{Watkins1989} combined with deep neural networks \cite{LeCun2015} and experience replay \cite{Lin1993b} to choose at every step one action from an action set which in our case represents governance management options. In the Q-learning box, the representation of the target function is depicted. To visualize Deep Q-Networks' functionality, we show a scheme for the function approximator via a deep neural network. In the experience replay box, the dot as the index of the observations in the replay buffer of size $ k $ represents an arbitrary time point.}
    \label{fig:agentenvironmentdqnlearner}
\end{figure*}

A scheme of how this framework is implemented is shown in Fig.\  \ref{fig:agentenvironmentdqnlearner}. In the following paragraphs, we provide more details on how we map the required parts for an MDP (i.e., concrete states in the environment, actions and reward function) to World-Earth models. We conclude this section with some technical notes about implementation and hyper-parameter search.  

\paragraph{Environment 1: The AYS model}
This model is a low-complexity model in three dimensions studied in \cite{heitzig2018} and described in more detail by \citeauthor{Kittel2017a} \cite{Kittel2017a}. It includes parts of climate change, welfare growth and energy transformation. As compared to classical Earth system models the AYS model is adapted. For simplicity carbon dynamics $ A $ is not modeled in an explicit carbon cycle but assumed to follow an exponential relaxation towards equilibrium. The relation of the wealth of a society is modeled through the economic output $ Y $, where the economy is assumed to have a constant basic growth rate.  A renewable energy source with learning by doing dynamics is implemented via a renewable energy knowledge stock $ S $. The state the agent observes at time $ t $ is therefore given by the tuple $ s_t = (A,Y,S)_t $, consisting of three numerical values. 
As sustainability boundaries we use a planetary boundary $A > A_{PB} = 345$\,GtC
and a social foundation boundary $Y > Y_{SF} = 4\cdot10^{13}$\,\$/yr.
For details, we refer to the Appendix or \cite{Kittel2017a}.

\paragraph{Environment 2: The copan:GLOBAL model} 
This model, studied by \citeauthor{Nitzbon2017} \cite{Nitzbon2017}, is a conceptual model that describes the coevolution of natural and economic subsystems of the Earth.  The model is meant for qualitative understanding of the complex interrelations rather than for quantitative predictions. Climate is represented as a global carbon cycle involving stocks of terrestrial carbon $ L $, atmospheric carbon $ A $ and geological carbon $ G $, which influence the global mean temperature $ T $. On the other hand socio-economic concepts, expressed through population $ P $ with capital $ K $, are used to describe the flows of biomass and fossil fuels between society and nature. In \cite{Nitzbon2017}, the authors consider a scenario where renewable energy does not exist. We extend the model for this study by including renewable energy use via a learning-by-doing dynamics for the renewable energy knowledge stock $ S $, in the same fashion as it was done in \cite{Donges2018} for a regionalized version of \cite{Nitzbon2017}. The state $ s_t $ is thus determined by the tuple $ s_t=(L,A,G,T,P,K,S)_t $. 
Similarly, we use again $A > A_{PB} = 345$\,GtC
and a social foundation boundary for consumption of $W > W_{SF} = 7850$\,\$/yr per-capita as sustainability boundaries. For details of the system dynamics the reader is refereed to the Appendix or \cite{Nitzbon2017,Donges2018}.

\paragraph{Action set} The action set $ \mathcal{A} $ represents certain governance management options. It consists of no extra management (called default), a carbon tax, subsidies of renewable energies, for the c:GLOBAL environment of a nature protection policy and all possible combination of these management options. Depending on the specific environment, each action alters the dynamics of the state variables. For details, we refer to the Appendix. 

\paragraph{Reward function} 
Reward functions express the agent's preferences over state-action pairs and therefore control the learning process. Since we implemented our own environments and are only interested in keeping them within certain bounds, we have freedom of choice for the reward function. The very simple reward functions we used are the following:
\begin{itemize}
    \item \textit{Survival reward}: provide a reward of $ 1 $ if the state $ s_t $ is the within the boundaries, else 0. 
    \item \textit{Boundary distance reward}: calculate the distance of the state $s_t$ to the boundaries in units of distances from the current state of the Earth to the boundaries. This distance is provided as a reward.
\end{itemize}
Depending on the chosen reward function, the trajectories found by the agent differ. In the case of survival reward, the agent is only interested in staying within the boundaries, whereas in the latter case of the boundary distance reward the agent tries to detect trajectories resulting in a large distance to the boundaries. 

\paragraph{Implementation}
After the experience replay memory is filled with experiences from an agent that acts randomly in the environment, the learning process runs as follows. The agent is trained for a fixed number of episodes. A start position within the boundaries is randomly drawn from a uniform distribution of states around the current state. The number of iteration steps during one single learning episode is limited to a maximum of $ T $. The end of one learning episode is determined either when $ T $ is reached or ended prematurely at time $ t $ either when a boundary is crossed or when approximate convergence to a fixed point is detected. In the latter case, the remaining future rewards are estimated with a discounted reward sum for the remaining time $ T-t $ of the reward $ r_t $. In any case, after the end of a learning episode, the environment is reset to time $ t=t_0 $ and a new start point $ s_{t_0} $ within the boundaries of the environment is randomly drawn. 

\paragraph{Hyper-parameter tuning}
For each environment, we trained a different network. We systematically investigated the effect of various parameters for the learning success, such as the discount factor, the training data set size or the exploration-exploitation trade-off to get an optimal hyper-parameter set for every environment. The exploration rate $ \epsilon $ starts with $ 1 $ and decays over time. We achieved the best performance for a replay buffer (i.e., the memory size) of $ 10^5 $ which is less than the default value in many DRL algorithms (e.g., \cite{Mnih2013, Wang2015, Hessel2018}) but in accordance with the work of \cite{Zhang2017} stating that the size of the replay buffer is crucially environment-dependent and needs a careful tuning. A full list of all hyper-parameters can be found in Table \ref{tab:List_hyperparameters} in the Appendix. 

The neural network is based on the following architecture. The input layer of the size equalling the dimension of the state space is followed by two fully-connected hidden layers each one consisting of 256 units. The output layer is a fully connected linear layer that provides an output value for each possible action in the action set, representing the estimated Q-value of that action for the state given by the inputs. For minimizing the loss function, instead of simple stochastic gradient descent (SGD) the Adam optimizer \cite{Kingma2014} is used due to its better performance than SGD in DRL applications as reported in \cite{Hessel2018}.

\section{Application to World-Earth Models}
Based on our proposal outlined above, we implemented an agent that learns by using a DRL (see section \ref{sec:drl}) to manage the environments described in Sect.\ \ref{sec:aei}. The agent is trained for a maximum number of $ 10^4 $ episodes, where the learning success is evaluated every $ 50 $ episodes. Single simulation experiments can be carried out on standard notebook computers in a reasonable computing time (one to two hours on a single machine). 
Using a tuned hyper-parameter set (see Table \ref{tab:List_hyperparameters} in the Appendix for details), we can formulate three key findings of this work that is outlined below. First, we find that learning in terms of increasing rewards in the environments is indeed possible. Second, we investigate the specific pathways found by the learner and observe that the agent acts with great farsightedness. Moreover, we see a general strategy behind the detected trajectories that the learner has developed. Third, we explore that the agent also achieves good performance in scenarios in which the state space is only partially observable to the agent.

\subsection{Training and Stability}
In order to verify the overall applicability of our algorithm, we first analyze the learning behavior in general. Unlike in supervised learning, where one can evaluate the performance of an algorithm by evaluating it on a set of test data, it is not obvious how to evaluate accurately the training progress an agent makes in RL problems. Here, we stick to the method used by \citeauthor{Mnih2013} \cite{Mnih2013} visualizing the training properly. We plot the total reward the agent collects during one run over the number of learning episodes. Each value is computed as a running average over 50 episodes. Each curve is the average of 100 independent simulations.

\begin{figure}
    \centering
    \includegraphics[width=1.\linewidth]{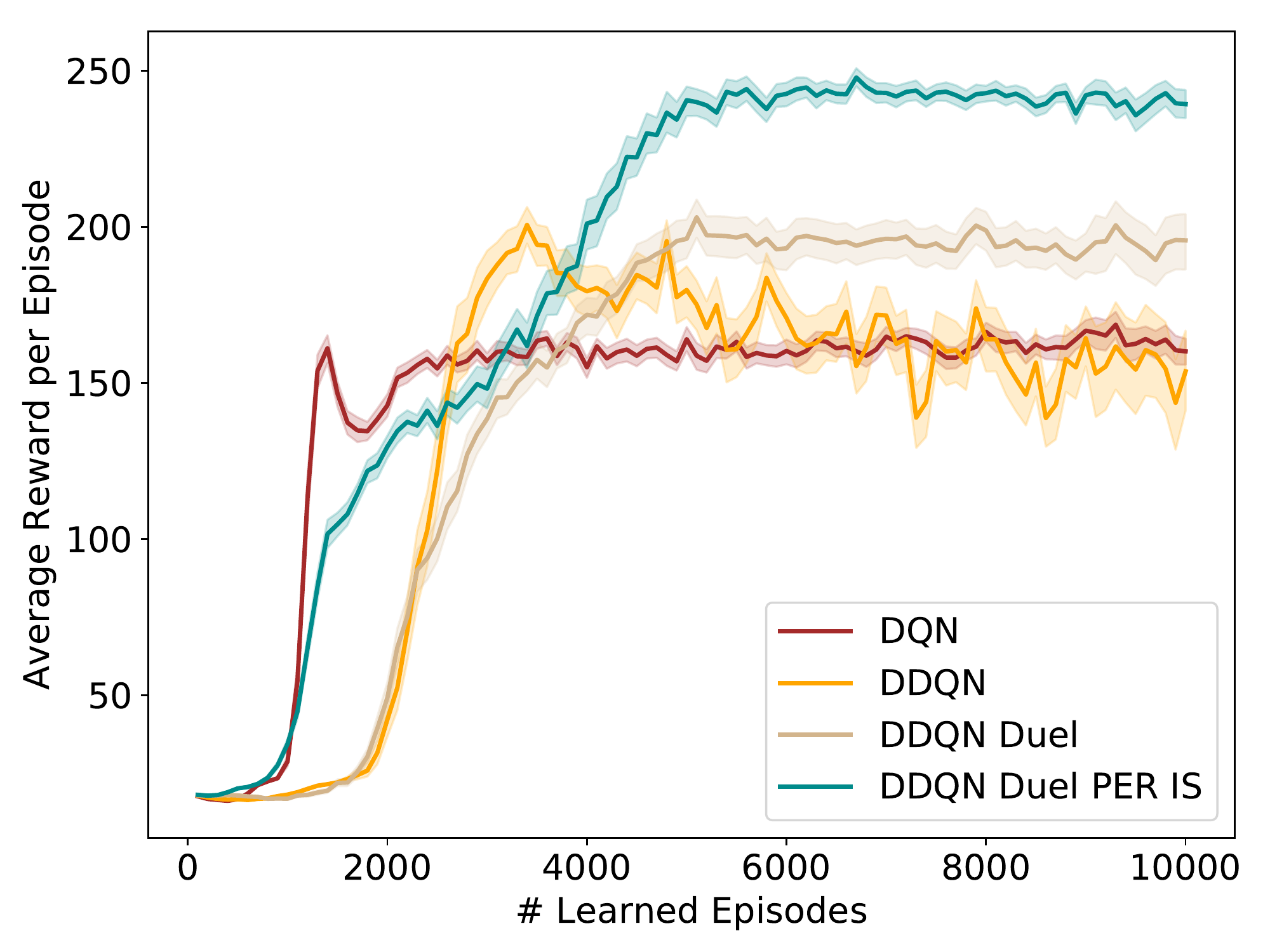}
    \caption{Development of total average reward per episode. The average reward is a running average over the last 50 episodes of the sum of all rewards gained during one training episode. The curves are an average of $ 100 $ independent simulations of the AYS-model. The light bands show $ 95\% $ confidence intervals for the expected values estimated by these averages.  Different Deep-Q-Network architectures are analyzed: DQN=Deep Q-Networks, DDQN=Double DQN, DDQN Duel=Dueling network architecture with DDQN, DDQN Duel PER IS = DDQN Duel using prioritized experience replay together with importance sampling. The simulations were performed with a $\epsilon$-greedy policy with $\epsilon$ decaying exponentially from 1 to 0.01 at a decay rate of $\lambda=0.001$.}
    \label{fig:compare_learners}
\end{figure}

As a result, we see that after a certain number of episodes the average reward per episode significantly increases in our environments (see Fig.\ \ref{fig:compare_learners}). Obviously, the agent finds trajectories that reveal a high reward. In other words, it learns to manage the environment. We conclude that management can indeed be learned by the agent.

Furthermore, we observe that the learning of the agent is stabilized by using the extensions to DQN-Learning as outlined in \ref{sec:extensionsDQN}. The plot suggests that the usage of dueling network architectures combined with double-Q-learning (DDQN + Duel) and prioritized experience replay with importance sampling (PER IS) significantly increases the performance of our DQN-Agent. This is in good agreement with the results in \cite{Wang2015}. Therefore, all results outlined below are achieved by using our best performing agent (DDQN + Duel + PER IS), if not stated otherwise.
 Moreover, this is in qualitatively good accordance with other comparisons of different learning architectures, as, e.g., presented in \cite{Hessel2018} and the learning curves show a similar shape as seen in other DRL applications \cite{Mnih2013, Mnih2015, Hessel2018}.

\subsection{Management Pathways in World-Earth System Models}
In the following, we discuss the pathways in the two environments described in section \ref{sec:aei} that were found by using the outlined framework of DRL. In \ref{sec:pathways_ays} we explore the AYS model, in \ref{sec:pathways_global} the copan:GLOBAL model. Specifically, in both environments, we are interested whether the learner finds trajectories towards regions which we can associate with a safe and just operating space without violating sustainability boundaries. First, we present some successful examples. As a next step, we look at the specific trajectories in more detail, hoping to understand the general strategy the agent found to reach its aim (i.e., maximize the total reward).

\subsubsection{Pathways in the AYS-model} \label{sec:pathways_ays}
\begin{figure}
    \centering
    \includegraphics[width=1.\linewidth]{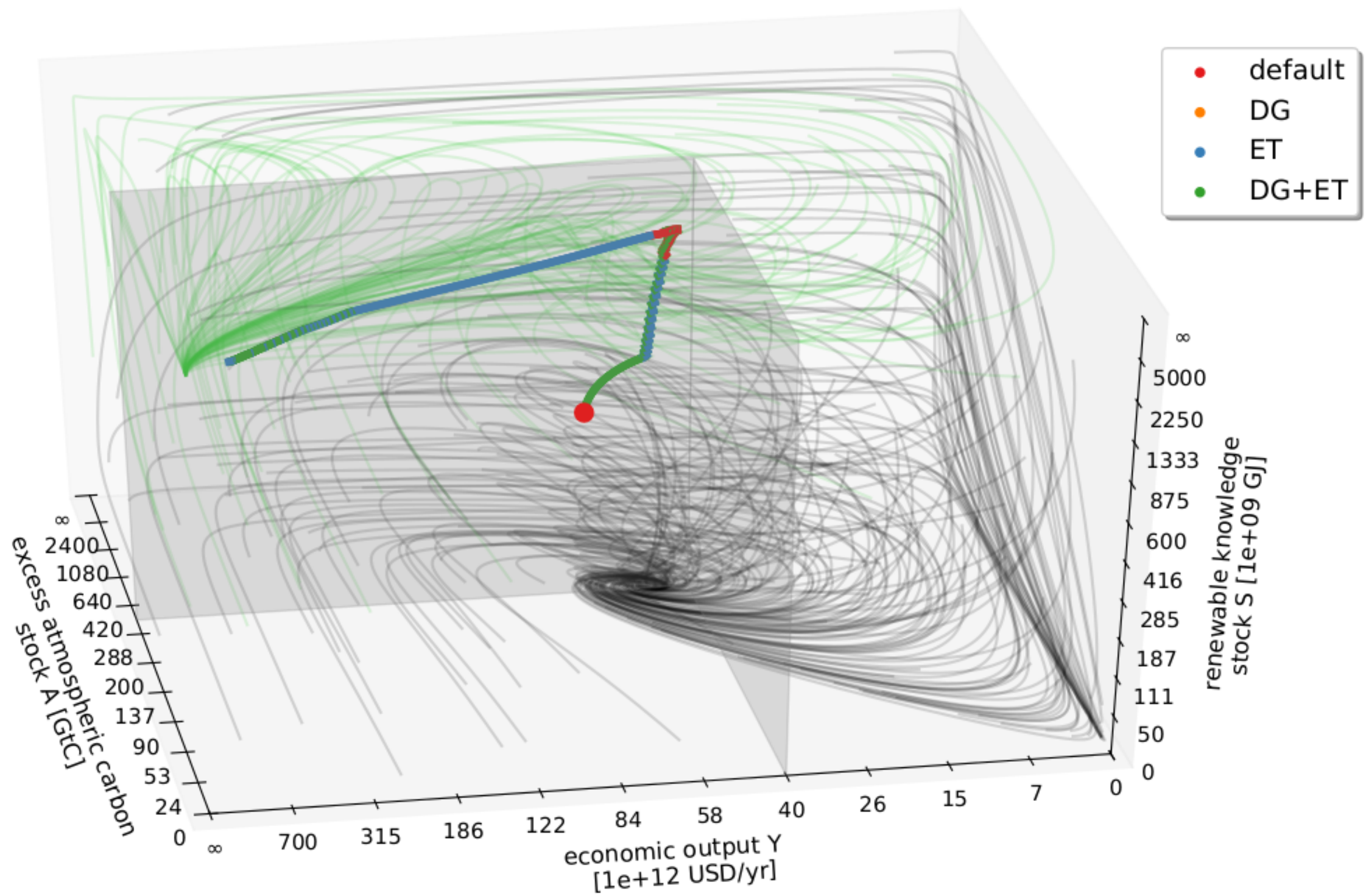}
    \caption{Dynamics of a stylized World-Earth system according to \citeauthor{Kittel2017a} \cite{Kittel2017a}. The default flow of the AYS model is sampled with thin trajectories with randomly distributed initial conditions. We used non-linearly scaled axes to account for the full space $ \mathbb{R}^3 $. Red dot in the center: Estimated current state of the world. Green lines: attraction basin of sustainable fix point which can be understood as the safe and just operating space. Black lines: attraction basin of the carbon-based economy. Grey surfaces: Sustainability Boundaries. In color: Example trajectory from the current state into a green future. The possible management options of the action set are: DG: Degrowth, and Energy-Transformation, i.e., Carbon tax + subsidies on renewables.}
    \label{fig:ays_plot}
\end{figure}

In the AYS model, the agent can choose between the following actions: ``energy transformation'' (taxing carbon emissions and/or subsidizing renewables) or ``degrowth management'' (reducing the basic economic growth rate) or neither or both of them. As a first analysis step, we look at the pathways the agent takes after it was trained for a sufficiently long time (i.e., the convergence of the learning is reached, see Fig.\ \ref{fig:compare_learners}). We find that even though the dynamics of the environment is unknown to the agent in advance, it is able to find trajectories within sustainability boundaries (see Fig.\ \ref{fig:ays_plot}) that were deemed impossible in another study based on a viability theory algorithm that used state space discretization \cite{Kittel2017a}.

Due to the setup of our framework, each of the two management options can only be switched on and off. In Fig.\ \ref{fig:ays_plot}, in the region near to the boundaries, the energy transformation (ET) option (representing an energy tax or subsidy) is switched on and off in short alternations, achieving essentially the effect that a continuous application of a smaller tax/subsidy would have. Hence, offering different tax levels as individual options might improve the learning success further. 

\begin{figure}
    \centering
    \includegraphics[width=1.\linewidth]{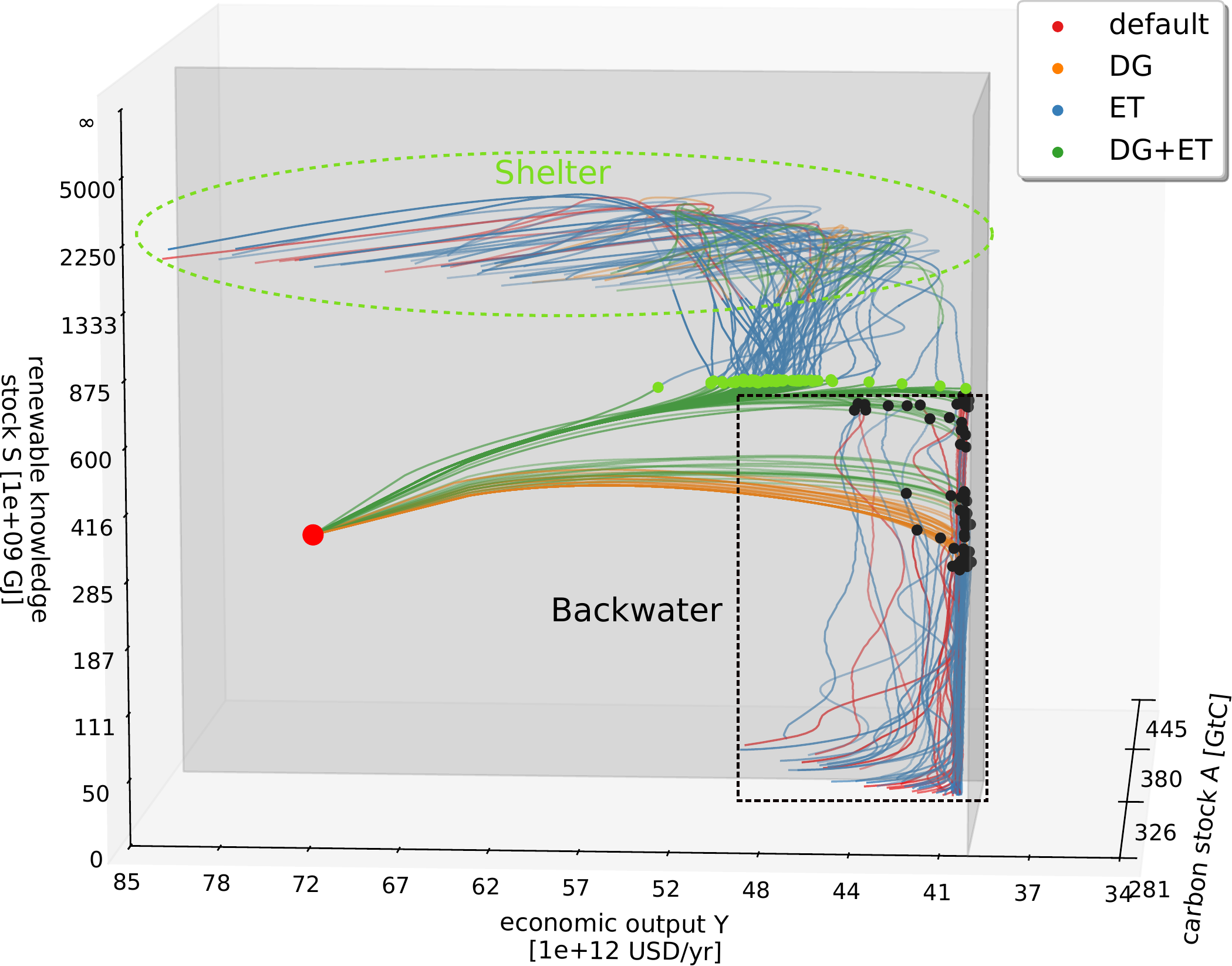}
    \caption{Analysis of predominant management strategies in 200 independent simulations that find a trajectory inside the boundaries (grey surfaces). Half of them use the boundary distance reward and go towards the green fixpoint within the "shelter" region where management can be stopped (green dashed line). The others use the survival reward and go into a fossil-based future within the "backwaters" region from which no return to the shelter region is possible (black dashed line). Management options: DG: degrowth, reducing the basic growth rate of the economy; ET: energy transformation, taxing carbon and/or subsidizing renewables. 
    Dots denote points of strongest gradient on each trajectory (green for going into the shelter, black for going into the backwaters). Here the predominant learning strategy changes as well. 
    The color of the trajectories shows the predominant management option used in each state. One can see that close to the shelter, no specific management option is preferred since the choice becomes irrelevant.}
    \label{fig:bassin_plot}
\end{figure}

To get a deeper understanding of the found solutions, we take a closer look at the different trajectories that were detected by using the DRL framework. Depending on the chosen reward function, the paths found by the agent differ. If the boundary distance reward is chosen, after sufficiently long learning, the agents always finds a path towards the "green fixpoint" at $(A,Y,S)=(0,\infty,\infty)$ where the distance to the boundaries is maximized. For the survival reward, the agent is only interested in staying within the boundaries. Therefore, it finds pathways leading to the green fixpoint as well as pathways towards a region close to the boundaries with $S=0$ where it then manages to stay. Although many viable paths are found by the learner, we find that the learning strategies that were found can be generalized. We analyzed the management options the agent uses most on different parts of the trajectories. They are depicted in Fig.\ \ref{fig:bassin_plot}. These different regions of predominant management options are now used for the following discussion. The different regions colored in Fig.\ \ref{fig:bassin_plot} may be analyzed with respect to a mathematical theory of the qualitative topology of the state space of a dynamical system with management options and desirable states, called \textit{topology of sustainable management} (TSM) \cite{Heitzig2016}. Interestingly, these regions can be seen to correspond roughly to some concepts from the TSM-framework, in particular, the concept of "shelter" and  "backwaters". The approximate locations of these regions are depicted by dashed lines in Fig.\ \ref{fig:bassin_plot}.

We identify a general strategy the agent uses. Starting from the current state, we found that in order to stay within the boundaries forever, it is not sufficient to use only one single management option of energy transformation (ET) or degrowth (DG) in the beginning. Rather, both ET and DG have to be applied to ensure keeping the system within the sustainability boundaries also in future times. However, to stay above the social foundations' boundary, at a certain time only ET has to be applied predominantly, leading to more or less sharp "turns" in the trajectory. If $S$ is large enough at this point, the turn is "upwards" and after some time, a region is reached where every trajectory is now leading towards unlimited growth of economic output and renewable knowledge regardless of the chosen management option, so that management can be "stopped". In TSM, such a secure region is called a \textit{shelter}. But if $S$ is too small at the turning point, the turn is "downward" towards $S=0$, staying close to the social foundation boundary. In \cite{Kittel2017a}, it was shown that this leads to a region called the {\em backwaters,} from which the shelter could not be reached any longer, but one can still stay within the boundaries by managing over and over again.

Summarizing, the agent learns that the timing of the particular change of management is of crucial relevance. A general interpretation of the resulting pathways would be that ET, e.g. via taxing fossils, is highly important to ensure further development. However, to reach a secure state without violating the sustainability boundaries, a degrowth policy is needed for some time as well. 

\subsubsection{Pathways in the c:GLOBAL-model} \label{sec:pathways_global}
\begin{figure}
    \centering
    \includegraphics[width=1.\linewidth]{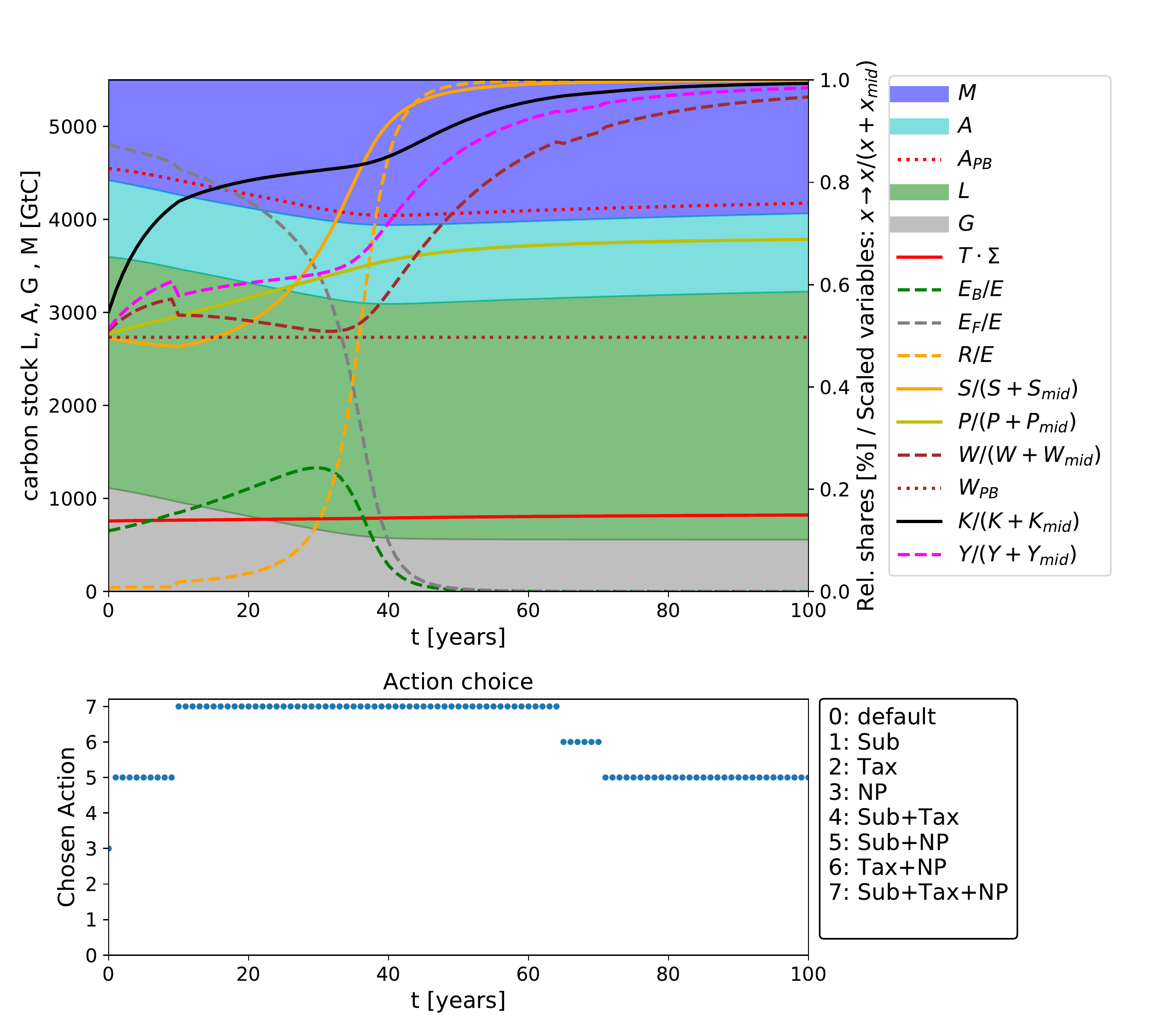}
    \caption{Exemplary trajectories for successful management in the model based on \cite{Nitzbon2017}. The upper graph shows the time evolution of the main variables, the lower the chosen actions at every timestep. Dynamical variables are displayed as colored bands and solid lines, derived variables as dashed lines, the planetary boundaries and social foundations in dotted lines. The total energy use is denoted as $ E=E_R+E_B+R $. For visual reasons we rescaled the $ S,P,W,K,Y $ with $ S_{mid}=5\cdot10^{11}\,\mathrm{bits} $, $ P_{mid}=6\cdot10^9\,\mathrm{H} $, $ W_{mid}=W_{SF}=7850\,\mathrm{\$/yr\,H} $, $ K_{mid}=5\cdot10^{13}\,\$ $ and $ Y=6.2\cdot10^{13}\,\mathrm{\$/yr} $. Since the system converges, only the first $ 100 $ years are shown. The available management options were Sub (subsidies on renewables), Tax (carbon tax on fossils), NP (nature protection for land use), and all combinations of these.}
    \label{fig:cgsuccessfulexampletrajectory}
\end{figure}

We verify that our framework works as well in higher-dimensional environments by applying it to the c:GLOBAL model. While classical approaches like viability theory are no longer well applicable because of the dimension, our DRL learner is also capable of detecting solutions towards a sustainable future in this model, see Fig.\ \ref{fig:cgsuccessfulexampletrajectory}. Here, one learning episode has a maximum length of $500\,\mathrm{yrs}$.  Successful trajectories often converge already after around $100\,\mathrm{yrs}$. However, to account for long-term effects, simulations were executed for times up to $500\,\mathrm{yrs}$ since we observed that seemingly converged trajectories sometimes transgressed boundaries at much later times, posing an additional challenge for the learner.  
The general strategy found by the learner turns out to be this. The NP option is used throughout and renewables are subsidized during most of the time. The crucial point is the timing of the carbon tax, which cannot be used immediately without violating the social foundation boundary. It is switched on only later and switched off again once renewables have passed through most of their learning curve.

An interesting observation regarding the farsightedness of the agent is the following. After some episodes, the agent often uses trajectories that do not use any management during the years $20$--$60$, which keeps the system within the boundaries for some time but leads to a violation of $ A_{PB} $ later for some $ t>100 \, \mathrm{yrs} $. Only after many more episodes, the agent learns to act with foresight and use management options early on that only make a recognizable difference much later and avoid crossing the boundaries. This is indeed a key feature for the success of DRL and shows the potential power of the method. One example trajectory can be found in Fig.\ \ref{fig:cgnonsuccessfulexampletrajectory} in Appendix.

However, taking a look at the stability of the learning (see Fig. \ref{fig:compare_part_obs_state}), we observe that the learning success in the copan:GLOBAL model also decreases again after a still larger number of episodes. As a possible explanation, we suggest that this is connected to the replay buffer. To avoid this phenomenon, the replay buffer needs to contain experiences especially about the timesteps where the dynamics of the system changes significantly \cite{Zhang2017}. After many successful runs, we still continue collecting observations in the memory buffer at every timestep. Therefore, it mostly contains experiences for time points $t>50\,\mathrm{ys} $. However, especially the first timesteps are crucial to avoid transgressing boundaries at later times as outlined above. These are therefore essential for the learning success. It seems that the agent tends to forget about experiences from early timesteps and the learning success decreases. Further investigation considering the question which experiences should be stored in the replay buffer could be a first step to overcome this issue.

\subsection{Partial Observability and Noise}
\begin{figure}
    \centering
    \includegraphics[width=1.\linewidth]{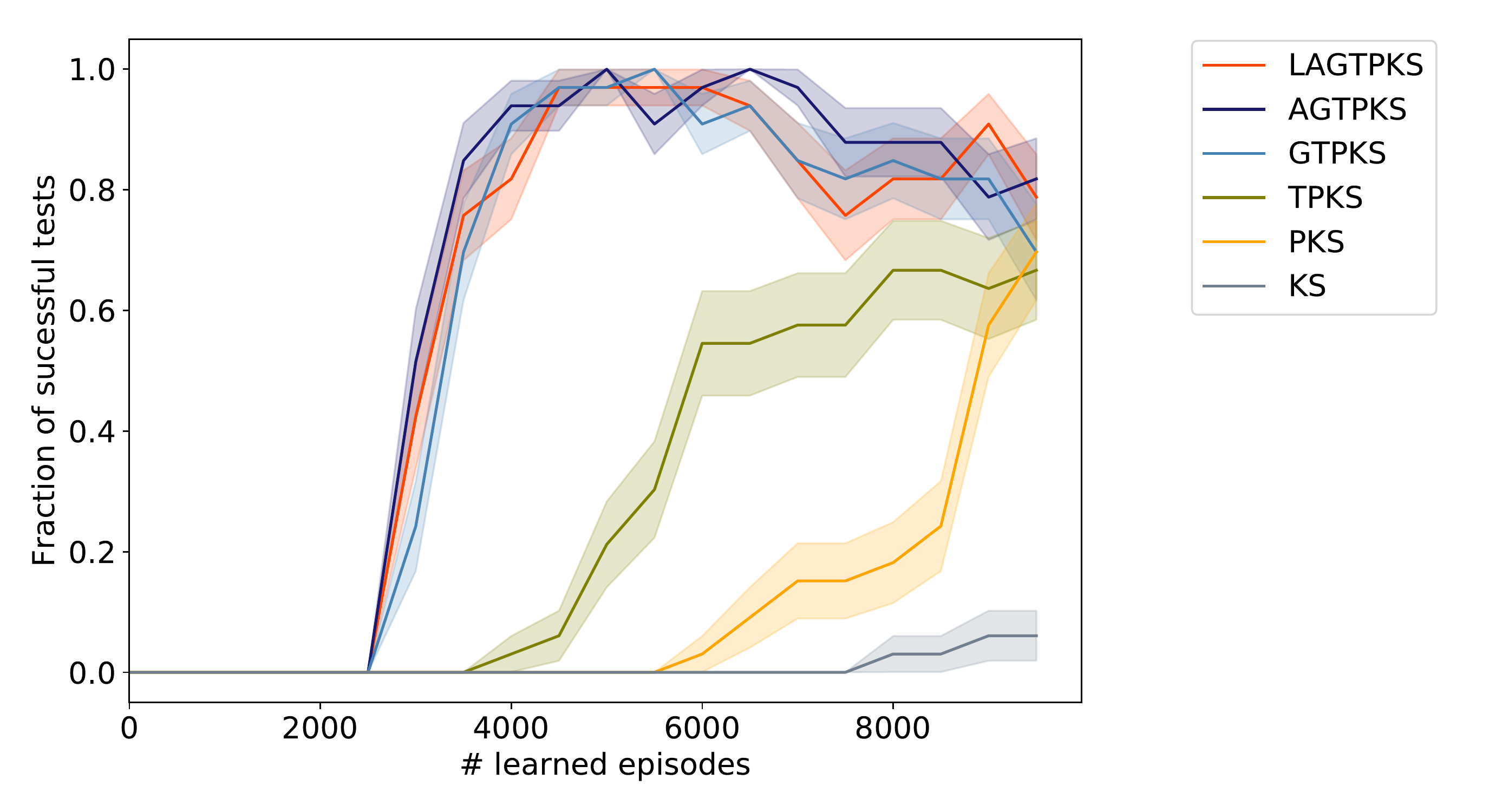}
    \caption{Percentage of tests the agent passes successfully given different information about the state. An episode is considered successful if the agent, starting from the current state, manages to reach the shelter region where management can be turned off.
    For each set of dynamical variables, we simulated 100 independent learning processes and show 95\% confidence bands for the reported percentages as estimates of the true success probabilities. The legend lists the variables observable by the agent: L = terrestrial carbon, A = atmospheric carbon, G = geological carbon, T = global mean temperature, P = population, K = capital, S = renewable energy knowledge stock.}
    \label{fig:compare_part_obs_state}
\end{figure}

As a generalization of Markov decision processes, partially observable Markov decision processes (POMDP) are of great research interest. Here, the agent is only able to observe only part of the actual system state \cite{Spaan2012}. 
We are interested in the performance of our DRL agent under such observational constraints since a real-world manager will only have access to vastly restricted information about the Earth system's current state. Moreover, we added noise to the observations of the agent. Our experiments show (see Fig. \ref{fig:compare_part_obs_state}) that even under partial observability of the state, the agent is still capable of detecting sustainable solutions. 
We observe that the learning curves for observing either the full state $(L,A,G,T,P,K,S)$ or only the variable combinations $(A,G,T,P,K,S)$ or $(G,T,P,K,S)$ have very similar shape. So it seems that there is little added value in observing the carbon stocks $L$ and $A$ when already observing the geological stock $G$ whose decline is essential for the timing of the carbon tax (but which is also the hardest to observe in reality).
However, even if we limit the agent's observation capabilities to the socio-economic variables $(P,K,S)$ the agent achieves a similar performance after a certain number of episodes, only considerably later. This can be explained by the dominant force humans exert on the Earth system.

To test the robustness of the DRL algorithm for a noisy state input, we added white observational noise on the input state $s_t$ the agent receives from the environment. Not surprisingly, noise can disturb the agent's learning and lead to a massive decrease in performance if the environment gets more complex. See Fig. \ref{fig:noise_strength} in Appendix for details. 
Neural networks are known to be vulnerable by perturbed input  \cite{Szegedy2013, Papernot2016}  and the harmful effect of noise has already been observed and discussed as well in DRL applications \cite{Behzadan2017, Behzadan2017a, Huang2017}. Still, for further experiments with more realistic scenarios, the influence of noise has to be investigated more systematically.

For the analysis of trajectories in the Earth system, we can deduce the following. Even if the full state will not be observable to the agent, it is just based on the distance boundary reward signal still able to sufficiently ``understand" the system's dynamics in order to find appropriate management pathways. Furthermore, in our experiments, we see that noise will be a limiting factor for some DRL algorithms. In simulations with very noisy environments, some preprocessing of the input state might be necessary to use DRL successfully.

\section{Conclusion}
 The main contribution of this work is the development of a framework for using DRL in Earth system models, mathematically formalized in a Markov decision process. Throughout this paper, we have combined the technique of deep reinforcement learning with Earth system modeling in order to detect global sustainable management strategies. We have presented a prototype for which we hope extensions based on our work will become a helpful tool to discover and analyze management pathways and to get a deeper understanding of the impact of global governance policies.

As a proof of concept, we have applied it to two exemplary models from Earth system science, taken from the World- Earth modeling literature. They include components of Earth system modeling as well as constraints of planetary boundaries and social foundations. We have shown that our algorithm successfully identified trajectories towards a secure region for the Earth system which a competing approach using viability theory and a discretization of the state space was not able to find \cite{Kittel2017a}.  Even very simple reward functions were sufficient and only partial observations of the system state were necessary for the learner to understand the complex, non-linear system's dynamics. However, noisy observations have presented a challenge. We have found significant learning improvements by using the combination of DQN with dueling network architectures and prioritized experience replay and importance sampling.

With respect to management strategies that the learner found in the AYS and the c:GLOBAL model, we can support the intuition that neither there is one single way for staying within the boundaries nor can the impact of global management be observed immediately. Rather we conclude from our models that only an intelligent combination and timing of global policies may lead to a sustainable future.
We found that besides making renewables more attractive, also a temporary slowing down of economic growth might be necessary for staying within planetary boundaries. 

Moreover, we have shown that our method is applicable as well in environments with only partially observable state spaces. Due to its connection to real-world problems \cite{Spaan2012}, for example in 3D navigation \cite{Mirowski2016}, partial observability of state spaces is widely discussed in the reinforcement learning community. Hence, in future work, the effects of reducing the dimensionality of the state space in our World-Earth system models need to be studied in more detail.

We used DRL to identify trajectories under certain constraints. Formally, this can be regarded as an optimization problem, which could be approached with other methods as well. E.g., the IAM community typically uses commercial solvers for the optimization of long-term social welfare functions which are influenced by nonlinear underlying dynamics. However, the choice of the welfare function is not directly intuitive and hard to justify straightforwardly \cite{Pindyck2017}. As an example, \citeauthor{Pindyck2017} \cite{Pindyck2017} puts forward the significant differences in the outcome of two established models in IAM. The results in \cite{Nordhaus2011} and \cite{Stern2007} differ widely, mainly based on the different values of the discount rates for the choice of which no uniform theory exists. However, in our models, the constraints imposed by sustainability boundaries, as well as the choice of simple reward functions, could be argued to be easier to justify and to understand intuitively in some contexts.

We encourage the reader to apply our framework to his or her preferred models. Since we formulated our problem as an MDP, our approach is not restricted to deterministic environments but can be generalized to environments that include stochastic dynamics and agent-based components.
One could think about replacing the global society used in the models above by agent-based models of regionally distributed interacting societies. Following the model developed by \citeauthor{Wiedermann2015} \cite{Wiedermann2015, Barfuss2017} which is a stochastic environment based on an adaptive network model, could be a first step in this direction. On the other hand, the biophysical dynamics could be incorporated in more detail as well by using more complex global vegetation models such as LPJ \cite{Sitch2003}.

Further, an interesting next step could be to use DQN agents to represent major real-world agents such as governments in a multi-agent environment setting. Here, first experiments in simple grid worlds have already been performed to investigate sequential social dilemmas \cite{Leibo2017} and common-pool resource appropriation \cite{Perolat2017}. Connections to game theory in the climate context are conceivable as well \cite{Heitzig2011, Barfuss2019}.

Another approach that might be promising is to include \textit{model-based} RL in our framework. Regarding computation time, model-based RL tends to be much more efficient \cite{Nagabandi2018}. The key difference is that model-free methods act in the real environment in order to collect rewards and update the action-value functions accordingly.
In contrast, the agent in model-based methods uses RL to learn a model of the environment and then predicts the system dynamics in a second step. Once the model is learned, actions can be chosen by using optimal control theory. 
Especially as environments in World-Earth models are often based on a set of biophysical and socio-economic differential equations, this approach might be promising.
However, highly complex environments often cannot be learned perfectly, such that solutions of this method involve the risk of being suboptimal. A possible approach to overcome this issue are recently developed algorithms that aim to combine advantages of both methods in one algorithm \cite{Pong2018}. 

Another fruitful exchange could emerge between the field of Earth system analysis and the field of safe and beneficial AI \cite{Amodei2016}. For example, the important question of the latter field of how self-learning agents can safely explore an environment without pursuing catastrophic action directly translates to finding sustainable policies in Earth system analysis. Here as well, management strategies need to navigate uncertain environments without activating tipping elements in the Earth system with potentially catastrophic impacts on human societies \cite{Lenton2008,Schellnhuber2009}.

\section*{Acknowledgements}

This work was developed in the context of the COPAN collaboration at the Potsdam Institute for Climate Impact Research (PIK). The authors thank the COPAN group for helpful discussions and comments. Moreover, FMS thanks the participants and organizers of the Eastern European Machine Learning (EEML) Summerschool for inspiring suggestions to this work. We are grateful for financial support by the European Research Council (via the ERC advanced grant project ERA), the Stordalen
Foundation (via the Planetary Boundaries Research Network PB.net), the Earth League's EarthDoc program and the Leibniz Association (project DOMINOES). 
The authors gratefully acknowledge the European Regional Development Fund (ERDF), the German Federal Ministry of Education and Research and the Land Brandenburg for providing resources on the high-performance computer system at PIK.

\section*{Appendix}

\subsection*{The AYS model environment}

In this environment, the observable state is composed of three real-valued components, the excess atmospheric carbon stock over pre-industrial levels $A\ge 0$ [GtC], the gross world economic product $Y\ge 0$ [\$/yr], and the global knowledge stock for producing renewable energy $S\ge 0$ [GJ].
The time evolution of these is given by three ordinary differential equations in which several additional, derived quantities occur that the agent cannot directly observe.
These auxiliary variables are total world demand for primary energy $U$ [GJ/yr], a relative price level of renewables $G$, resulting in a division of $U$ into renewable energy production $R$ and a flow of fossil energy $F$, and finally the global greenhouse gas emissions flow $E$ [GtC/yr] resulting from $F$.
We assume each unit of output $Y$ requires a fixed amount $1/\epsilon$ of energy and the two energy sources are used in proportion of relative price (see \cite{Kittel2017a} for a justification), so that 
\begin{align}
U &= Y/\epsilon, & F &= G U, & R &= (1 - G) U, & E &= F/\phi.
\end{align}
We assume the absolute price of fossils to remain constant and that of renewable energy to depend on the renewable knowledge stock in a power law relationship, so that the relative price of renewables vs.\ fossils has the form
\begin{align}
G &= \frac{1}{1 + (S/\sigma) \rho}.
\end{align}
Instead of assuming a carbon cycle as in the c:GLOBAL model (see below), we here simply assume atmospheric carbon stock declines exponentially towards its equilibrium value where excess atmospheric carbon vanishes, so that 
\begin{align*}
dA/dt &= E - A/\tau_A.
\end{align*}
Likewise, instead of assuming a classical economic growth model as in c:GLOBAL, we here simply assume gross world product grows at a fixed basic rate which is reduced in proportion to $A$ (interpreted as a proxy for climate damages),  
\begin{align*}
dY/dt &= (\beta - \theta A) Y.
\end{align*}
Finally, learning-by-doing makes the renewable knowledge stock grow with renewable energy production, and forgetting makes it decline exponentially:
\begin{align*}
dS/dt &= R - S/\tau_S.
\end{align*}
We use the following initial conditions and parameter estimates from \cite{Kittel2017a}:
energy efficiency $\epsilon = 147$ \$/GJ,
fossil combustion need $\phi = 4.7\cdot10^{10} $\,GtC/GJ,
break-even level of renewables $\sigma = 4\cdot10^{12}$\,GJ,
learning-by-doing exponent $\rho = 2$,
characteristic time of natural carbon uptake $\tau_A = 50$\,yrs,
basic economic growth rate $\beta = 0.03$\,/yr,
climate damage coefficient $\theta = 8.57\cdot10^{-5}$\,/yr/GtC,
characteristic time of forgetting $\tau_S = 50$\,yrs,
initial values $A_0 = 840$\,GtC, 
$Y_0 = 7\cdot10^{13}$\,\$/yr, 
$S_0 = 5\cdot10^{11}$\,GJ.

The AYS environment is an interesting minimum-complexity toy model for sustainability science because one can represent both the climate change planetary boundary and a wellbeing social foundation boundary in it by studying whether $A$ may stay below some threshold 
$A_{PB} = 345$\,GtC
and $Y$ does not drop below some minimum value 
$Y_{SF} = 4\cdot10^{13}$\,\$/yr.
In this paper, we assume the agent that represents the world community will try to avoid that the system converges to a fixed point with $S = 0$, $A > \bar A$ and $Y < \bar Y$, e.g.\ by making it instead go to $A=0$ and $S, Y = \infty$ without violating the bounds.
To do so, she has the option ``DG'' to reduce the basic growth rate to $\bar\beta = \beta/2 $, 
the option ``ET'' to support an energy transition and lower the break-even point to $\bar\sigma = \sigma \cdot (1/2)^{\rho} $ by subsidizing renewables and/or taxing fossils, and she can also use neither or both of these options.

\subsection*{The c:GLOBAL model environment}

The model underlying this environment is of a similar type but more complex, having seven dynamic variables, of which the agent can observe different subsets in our experiments, as well as several additional unobserved auxiliary variables.

Here, terrestrial carbon stock changes due to temperature-dependent photosynthesis (1st term) and respiration (2nd term), and due to harvesting of biomass $B$,
\begin{align*}
dL/dt &= (l_0 - l_T T) \sqrt{A/\Sigma}\, L - (a_0 + a_T T) L - B.
\end{align*}
Absolute atmospheric carbon stock $A$ changes due to photosynthesis, respiration, combustion of harvested biomass ($=-dL/dt$), and ocean-atmosphere diffusion,
\begin{align*}
dA/dt &= - dL/dt + \delta (M - m A).
\end{align*}
Geological carbon stock $G$ declines because of extraction of fossil fuels $F$,
\begin{align*}
dG/dt &= - F.
\end{align*}
Global mean temperature converges to a value dependent on $A$ due to the greenhouse effect and is hence measured for simplicity on a nonlinear scale in units of atmospheric carbon per land surface area, so that 
\begin{align*}
dT/dt &= g (A/\Sigma - T).
\end{align*}
Population $P$ has a fertility (1st term) and mortality (2nd term) that depend on wellbeing $W$,
\begin{align*}
dP/dt &= P \left( \frac{2 W W_p}{W^2 + W_p^2}\,p - \frac{q}{W} \right).
\end{align*}
Physical capital $K$ grows since part of GWP $Y$ is invested, and decays exponentially,
\begin{align*}
dK/dt &= i Y - k K.
\end{align*}
For renewable knowledge stock $S$, we assume the same dynamics as in the AYS model, 
\begin{align*}
dS/dt &= s_R R - s_S S.
\end{align*}
Since total carbon is fixed at $C^\ast$, maritime carbon stock $M$ is
\begin{align*}
M &= C^\ast - L - A - G.
\end{align*}
Usage of the three assumed perfectly substitutable energy forms of biomass $B$, fossil $F$, and renewable energy flow $R$ is determined by a general price equilibrium model (see \cite{Nitzbon2017,Donges2018}) that leads to these equations:
\begin{align}
B &= \frac{a_B}{e_B}\frac{L^2 (P K)^{2/5}}{(a_B L^2 + a_F G^2 + a_R S^2)^{4/5}} \; ,        \label{eqn:B}\\
F &= \frac{a_F}{e_F}\frac{G^2 (P K)^{2/5}}{(a_B L^2 + a_F G^2 + a_R S^2)^{4/5}}\; ,        \label{eqn:F}\\
R &= a_R\frac{S^2 (P K)^{2/5}}{(a_B L^2 + a_F G^2 + a_R S^2)^{4/5}}        \label{eqn:R}.
\end{align}
Economic output is proportional to energy input,
\begin{align*}
Y &= y_E (e_B B + e_F F + R).
\end{align*}
Finally, wellbeing is determined by per-capita consumption and ecosystem services which are assumed proportional to terrestrial carbon density:
\begin{align*}
W &= \frac{(1 - i) Y}{P} + w_L\frac{L}{\Sigma}.
\end{align*}
We use the following initial conditions and parameter estimates from \cite{Nitzbon2017,Donges2018}, which are based on data from year 2000: 
initial values $L_0 = \SI{2480}{\carbon} $ (GtC=gigatons carbon), 
$A_0 = \SI{830}{\carbon}$, 
$G_0 = \SI{1125}{\carbon}$, 
$T_0 = \SI{5.05e-6}{\carbon\meter\inversesquare}$, (global mean surface air temperature is not measured in Kelvin but for simplicity in carbon-equivalent degrees, i.e. GtC),
$P_0 = \SI{6e9}{\humans}$ (H=humans), 
$K_0 =\SI{5e13}{\dollar} $, 
$S_0 = \SI{5e11}{\bits}$.

The parameters are: 
photosynthesis parameters $l_0 = \SI{26.4}{\kilo\meter\year\inverse\carbon}^{-1/2}$ and
$l_T = \SI{26.4}{\cubic\kilo\meter\year\inverse\carbon}^{-3/2}$,
total land mass $\Sigma = \SI{1.5e8}{\square\meter}$,
respiration parameters $a_0 = \SI{0.03}{\year\inverse}$ and $a_T = \SI{1.1e6}{\square\kilo\meter\year\inverse\carbon\inverse}$,
diffusion coefficient $\delta = \SI{0.01}{\year\inverse}$,
solubility coefficient $m = 1.5$,
strength of greenhouse effect $g = \SI{0.02}{\year\inverse}$,
peak fertility wellbeing level $W_p = \SI{2000}{\dollar\year\inverse\humans\inverse}$,
peak fertility $p = \SI{0.04}{\year\inverse}$,
mortality coefficient $q = \SI{20}{\dollar\year\inversesquare}$,
savings and capital depreciation rates $i = 0.25$ and $k = \SI{0.1}{\year\inverse}$,
knowledge accumulation $s_R = \SI{1.}{\bits\carbon\inverse}$ and 
forgetting parameters  $s_S = \SI{1/50}{\year\inverse}$,
total carbon $C^\ast = \SI{5500}{\carbon}$,
energy subsector productivities $a_B =1.5\cdot10^4  \text{GJ}^5\text{yr}^{-5}\text{GtC}^{-2}\text{\$}^{-2}\text{H}^{-2}$, 
$a_F = 2.7\cdot10^5  \text{GJ}^5\text{yr}^{-5}\text{GtC}^{-2}\text{\$}^{-2}\text{H}^{-2}$, 
$a_R = 9\cdot10^{-15}  \text{GJ}^5\text{yr}^{-5}\text{bits}^{-2}\text{\$}^{-2}\text{H}^{-2}$,
energy efficiencies $e_B = \SI{4e10}{\giga\joule\carbon\inverse}$, 
$e_F = \SI{4e10}{\giga\joule\carbon\inverse}$, 
final sector productivity $y_E = \SI{120}{\dollar\giga\joule\inverse}$,
wellbeing-sensitivity on ecosystem services $w_L = 0$.

\subsection*{Unsuccessful Management in c:GLOBAL}
\begin{figure}[H]
    \centering
    \includegraphics[width=1.\linewidth]{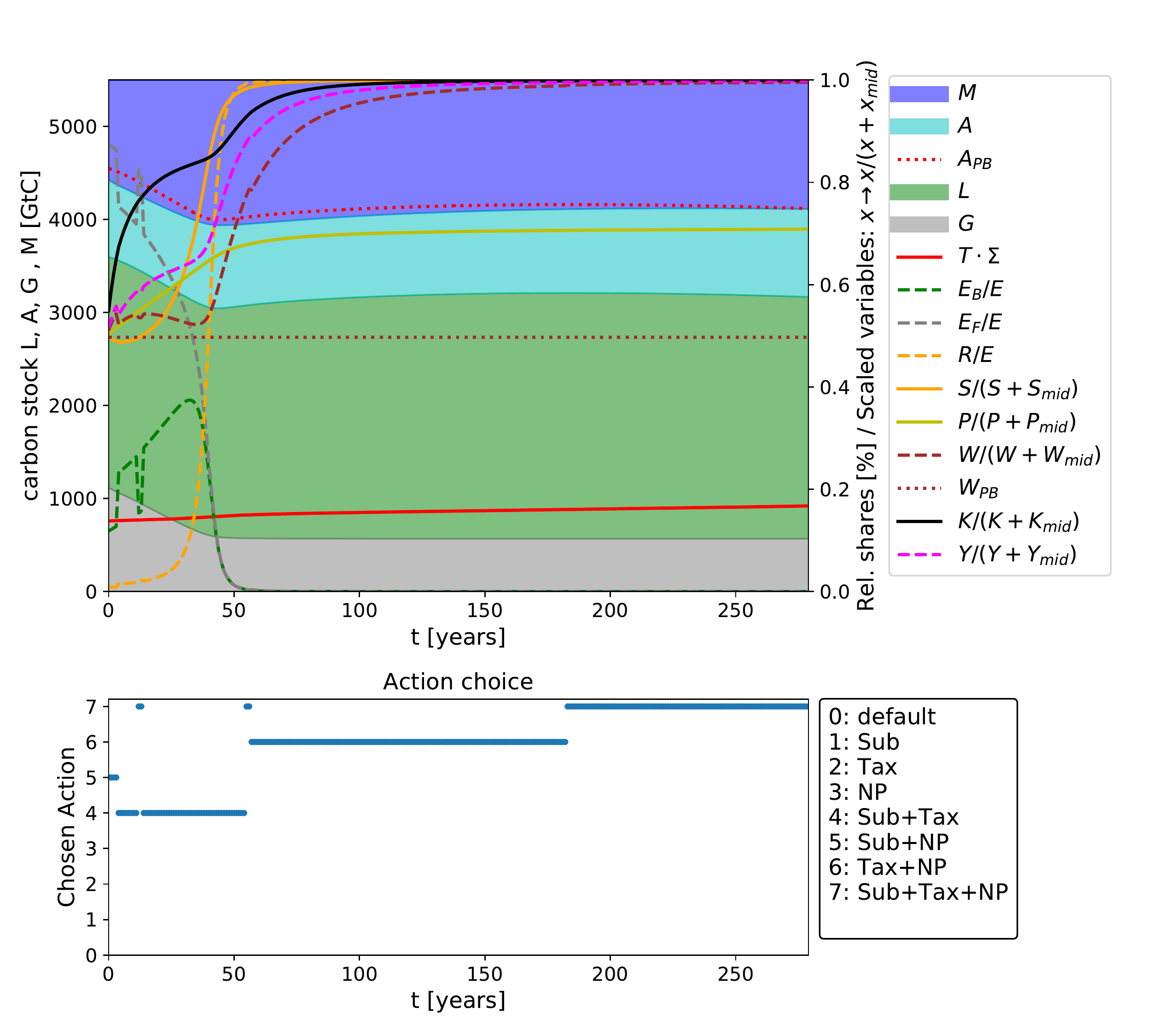}
    \caption{Exemplary trajectories for unsuccessful management in the model based on \cite{Nitzbon2017}. The upper graph shows the different trajectories, the lower the chosen action at that time. Dynamical variables are displayed in solid lines, derived variables in dashed lines, planetary boundaries and Social Foundations in dotted lines. The total energy use is denoted as $ E=E_R+E_B+R $. For visual reasons we rescaled the $ S,P,W,K,Y $ with $ S_{mid}=5\cdot10^{11}\,\mathrm{bits} $, $ P_{mid}=6\cdot10^9\,\mathrm{H} $, $ W_{mid}=7850\,\mathrm{\$/aH} $, $ K_{mid}=5\cdot10^{13}\,\$ $ and $ Y=6.2\cdot10^{13}\,\mathrm{\$/a} $. Since the system converges, only the first $ 100 \, \mathrm{ys} $ have been plotted. The available management options were: Sub=Subsidies on renewables, Tax=Carbon tax on fossils, NP=Nature protection for landuse.}
    \label{fig:cgnonsuccessfulexampletrajectory}
\end{figure}

\subsection*{Noisy input to environments}
\begin{figure}[H]
    \centering
    \includegraphics[width=1.\linewidth]{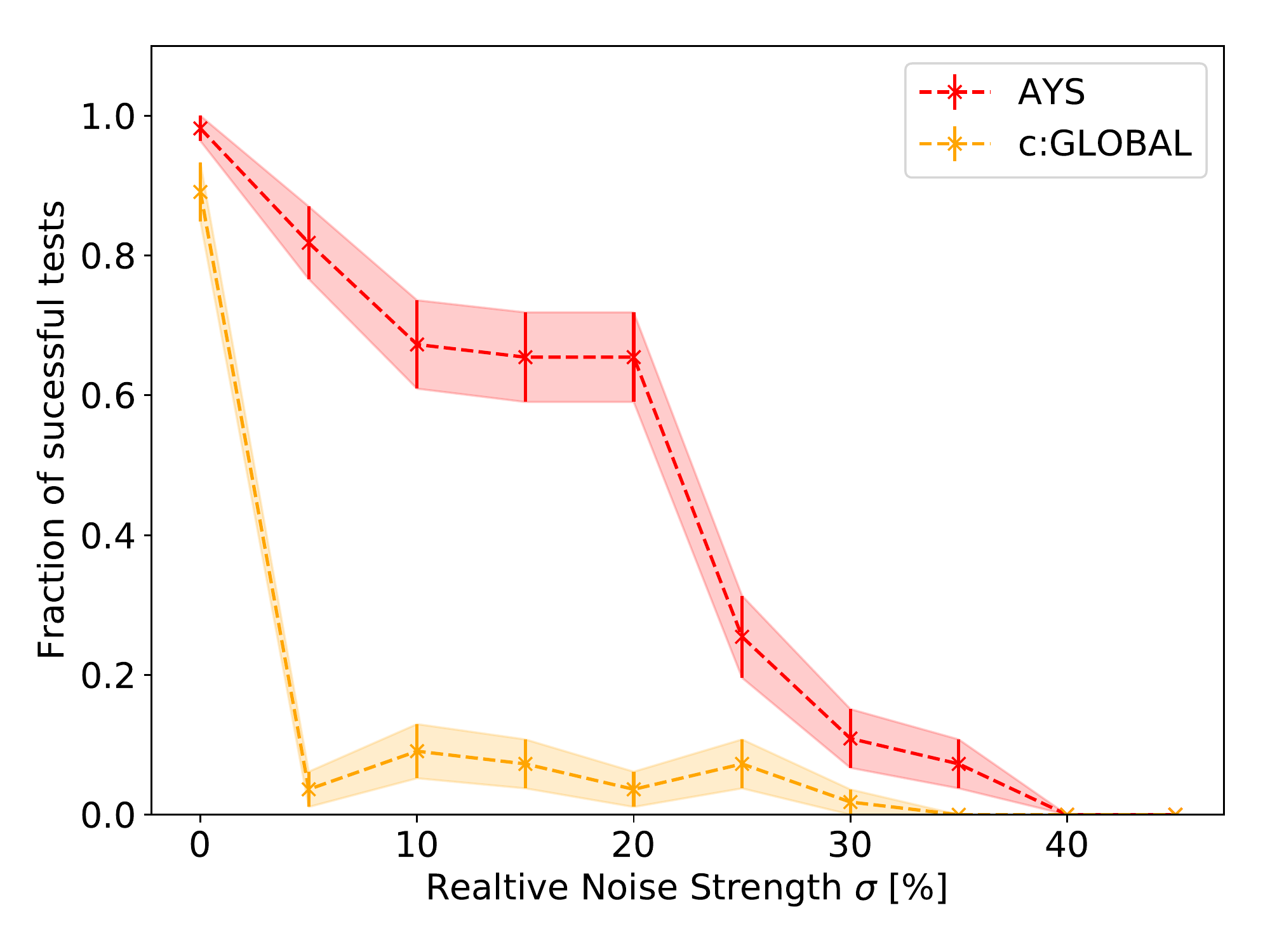}
    \caption{Percentage of successful tests for environments with different levels of noise strength $\sigma$. White noise is set on the input states, the strength of the noise could be up to $\sigma$ relative to the input state $s_t$. We compare the the learning success in the AYS as well as in the c:GLOBAL model. The agent is provided all dimensions of the state space. }
    \label{fig:noise_strength}
\end{figure}

\subsection*{List of hyper-parameters}
In Table \ref{tab:List_hyperparameters} the list of the hyper-parameters in AYS and c:Global environment is shown. 
The hyper-parameter search was mainly done based on own exploration. Due to high computational costs, no systematic grid search was performed, but as one parameter was tested the remaining were fixed at their previously explored optimal values. For the priority of transition $\alpha $, the initial importance sampling weighting $ \beta_0 $ and the Adam optimizer learning rate the recommended values in \cite{Schaul2015} and \cite{Hessel2018} were used. 
\begin{table*}[t]
    \centering
    \begin{tabular}{|l|c|c|l|}
        \hline
        \textbf{Hyperparameter}                      & \textbf{Value AYS} & \textbf{Value c:Global} & \textbf{Description}                                                                                                                                                               \\ \hline\hline
        batch size                                   & 32                 & 32                      & \begin{tabular}[c]{@{}l@{}}Number of training observations\\ over which Q-value function update is computed\end{tabular}                                                           \\ \hline
        replay memory size                           & $1\cdot 10^{5}$    & $1\cdot 10^{5}$         & Number of stored observations in replay memory.                                                                                                                                     \\ \hline
        initial exploration                          & 1                  & 1                       & Initial value in $\epsilon$-greedy policy.                                                                                                                                          \\ \hline
        final exploration                            & 0.01               & 0.001                   & Final value in $\epsilon$-greedy. policy                                                                                                                                            \\ \hline
        decay rate exploration                       & 0.001              & 0.001                   & Exponential decay of $\epsilon$ towards final value.                                                                                                                                \\ \hline
        target network update frequency              & 100                & 200                     & \begin{tabular}[c]{@{}l@{}}The number of episodes after which\\ the parameters of the target network \\ are updated to the current network parameters.\end{tabular}                 \\ \hline
        Adam learning rate                           & 0.00025            & 0.00025                 & The initial learning rate in Adam optimizer                                                                                                                                                \\ \hline
        discount factor $\gamma$                     & 0.96               & 0.96                    & Discount factor used in Q-learning. update                                                                                                                                          \\ \hline
        priority of transition $ \alpha $                       & 0.6                & 0.6                     & \begin{tabular}[c]{@{}l@{}}In Prioritized Experience Replay: The exponent \\ determines how much prioritization is used. \end{tabular}                                               \\ \hline
        initial importance sampling weight $\beta_0$ & 0.4                & 0.4                     & \begin{tabular}[c]{@{}l@{}}In importance sampling $\beta$ is annealed\\ from $\beta_0$  to 1, which means its affect \\ is more relevant at the end of the simulation.\end{tabular} \\ \hline
    \end{tabular}
    \caption{Hyper-parameters for the AYS and the c:Global environment.}
    \label{tab:List_hyperparameters}
\end{table*}

\newpage
\section*{References}
\bibliography{myBibFiles.bib}

\end{document}